\documentclass[aps,prx,twocolumn,superscriptaddress,longbibliography]{revtex4-2}
\usepackage{amsmath,amsthm,amsfonts}
\theoremstyle{definition}
\newtheorem*{proposition}{Proposition}

\usepackage{multirow}
\usepackage{array}

\usepackage{float}
\usepackage{graphicx}
\usepackage{afterpage}
\usepackage{caption}
\usepackage{subcaption}
\usepackage[font=small,labelsep=quad,format=plain]{caption}

\usepackage[colorlinks=true,citecolor=blue,linkcolor=blue,urlcolor=blue]{hyperref}
\usepackage{cleveref}
\crefname{equation}{Eq.}{Eqs.}
\crefname{section}{Sec.}{Secs.}
\crefname{subsection}{Sec.}{Secs.}
\crefname{appendix}{Appendix}{Appendices}
\crefname{figure}{Fig.}{Figs.}
\crefname{table}{Table}{Tables}
\crefname{proposition}{}{}
\crefname{corollary}{}{}
\usepackage{bm}
\usepackage{physics}
\input{Qcircuit}
\usepackage[dvipsnames]{xcolor} 

\newcommand{\Z}{\mathbb{Z}}
\newcommand{\E}{\text{E}}
\newcommand{\Var}{\text{Var}}

\begin{document}

\title{Quantum Computation of Finite-Temperature Static and Dynamical Properties of Spin Systems Using Quantum Imaginary Time Evolution}

\author{Shi-Ning Sun}
\affiliation{Division of Engineering and Applied Science, California Institute of Technology, Pasadena, CA 91125, USA}
\author{Mario Motta}
\affiliation{IBM Quantum, IBM Research Almaden, San Jose, CA 95120, USA}
\author{Ruslan N. Tazhigulov}
\affiliation{Division of Chemistry and Chemical Engineering, California Institute of Technology, Pasadena, CA 91125, USA}
\author{Adrian~T.~K.~Tan}
\affiliation{Division of Engineering and Applied Science, California Institute of Technology, Pasadena, CA 91125, USA}
\author{Garnet Kin-Lic Chan}
\email{garnetc@caltech.edu}
\affiliation{Division of Chemistry and Chemical Engineering, California Institute of Technology, Pasadena, CA 91125, USA}
\author{Austin J. Minnich}
\email{aminnich@caltech.edu}
\affiliation{Division of Engineering and Applied Science, California Institute of Technology, Pasadena, CA 91125, USA}
\date{\today}

\begin{abstract}

Developing scalable quantum algorithms to study finite-temperature physics of quantum many-body systems has attracted considerable interest due to recent advancements in quantum hardware. However, such algorithms in their present form require resources that exceed the capabilities of current quantum computers except for a limited range of system sizes and observables. Here, we report calculations of finite-temperature properties including energies, static and dynamical correlation functions, and excitation spectra of spin Hamiltonians with up to four sites on five-qubit IBM Quantum devices. These calculations are performed using the quantum imaginary time evolution (QITE) algorithm and made possible by several algorithmic improvements, including a method to exploit symmetries that reduces the quantum resources required by QITE, circuit optimization procedures to reduce circuit depth, and error mitigation techniques to improve the quality of raw hardware data. Our work demonstrates that the ansatz-independent QITE algorithm is capable of computing diverse finite-temperature observables on near-term quantum devices.

\end{abstract}
\maketitle

\section{Introduction}
  
Quantum computers have long been considered as a potential tool to simulate quantum many-body systems \cite{feynman_1982,lloyd_1996,georgescu_2014}. While near-term quantum devices have made rapid progress in simulating ground-state properties \cite{peruzzo_2014,omalley_2016,kandala_2017,colless_2018,kandala_2019,ma_2020,google_hartree-fock_2020} and dynamics \cite{islam_2013,zhang_2017,smith_2019,chiesa_2019,francis_2020}, the study of finite-temperature physics on quantum computers is less understood and established \cite{bauer_2020}. Early works on digital quantum simulation of finite-temperature physical systems involved thermalizing the quantum simulator by coupling to a bath comprised of ancilla qubits \cite{terhal_2000,poulin_2009,riera_2012} or sampling thermal states using quantum versions of the Metropolis algorithm \cite{temme_2011,yung_2012}. These schemes require prohibitively large numbers of qubits and deep circuits and are hence out of reach for near-term quantum devices. 

More practical variational algorithms have been proposed in recent years, such as protocols to construct thermofield double states \cite{martyn_2019,wu_2019} and machine learning-based methods to construct Gibbs states \cite{liu_2019,verdon_2019,chowdhury_2020,zoufal_2020,wang_2020}. However, the accuracies of these variational schemes depend on the quality of the ansatz. While other non-variational alternatives exist, they are subject to various assumptions. For example, the minimal effective Gibbs ansatz \cite{cohn_2020} algorithm generates a minimal ensemble of pure states but presumes the eigenstate thermalization hypothesis.

Recently, the quantum imaginary time evolution (QITE) algorithm was introduced \cite{motta_2019}. Compared to variational-based algorithms of imaginary time evolution on quantum computers \cite{mcardle_variational_2019,yuan_2019,beach_2019}, QITE is ansatz-independent. The QITE algorithm approximates imaginary time evolution with unitary operators over a domain of qubits and is able to reach the ground states of systems with a few sites. QITE can also be used to calculate finite-temperature quantities, for instance by combining with sampling techniques such as the minimal entangled typical thermal states (METTS) algorithm \cite{white_2009,stoudenmire_2010}, together denoted as the quantum METTS (QMETTS) algorithm. However, the original work on QITE \cite{motta_2019} focused on the general formalism, while reduction and optimization of quantum resources were not thoroughly investigated. Subsequent development of QITE \cite{yeter-aydeniz_practical_2020,nishi_2020,gomes_2020,yeter-aydeniz_scattering_2020} proposed several variations of the original algorithm, but the practical evaluation of finite-temperature properties on existing quantum devices remains largely unaddressed. 

Here, we report QITE-based calculations of finite-temperature static and dynamical properties of one-dimensional spin systems with up to four sites on five-qubit IBM Quantum devices. The computed observables include finite-temperature energies, static and dynamical correlation functions, and excitation spectra. These calculations are made possible by several algorithmic improvements. First, we exploit symmetries in the spin Hamiltonians to reduce Pauli strings in the QITE unitaries, thus reducing the overall required quantum resources. Second, circuit optimization procedures including gate decomposition and circuit recompilation are used to further reduce circuit depth. Third, error mitigation techniques, namely post-selection, readout error mitigation and phase-and-scale correction, are used to improve the quality of raw hardware data. Our work demonstrates that with efficient use of quantum resources and effective error mitigation strategies, the ansatz-independent QITE algorithm is capable of computing diverse finite-temperature observables on near-term quantum devices.

This paper is organized as follows. In \cref{sec:theory} we review the QITE algorithm and propose a quantum circuit to evaluate finite-temperature dynamical correlation functions. In \cref{sec:methods} we introduce the algorithmic improvements including Pauli string reduction, circuit optimization and error mitigation that enabled us to obtain accurate results from hardware. Section \ref{sec:results} presents the results of our two-site and four-site calculations. Finally, we conclude and suggest directions for future studies in \cref{sec:conclusion}.

\section{Theory} 
\label{sec:theory}

\subsection{Quantum imaginary time evolution (QITE)}
\label{sec:qite}

We begin by reviewing the QITE algorithm in the context of a general Trotterization scheme of the imaginary time propagator. Consider imaginary time evolution on $N$ qubits under a Hamiltonian $\hat{H} = \sum_{m=1}^M \hat{h}[m]$, where each $\hat{h}[m]$ acts on a local set of qubits. Since the local terms $\hat{h}[m]$ are not commutative, we need to Trotterize the imaginary time propagator $e^{-\beta\hat{H}}$ by grouping local terms $\hat{h}[m]$ into Trotter terms $\hat{H}[l]$ such that each $\hat{H}[l]$ is a sum of local terms $\hat{h}[m]$ and $\hat{H} = \sum_{l=1}^{L} \hat{H}[l]$. For example, for a two-local Hamiltonian where each local term $\hat{h}[m]$ acts on qubits $m-1$ and $m$, setting $L = 2, \hat{H}[1] = \sum_{m=1}^{\lceil M/2 \rceil} \hat{h}[2m-1]$ and $\hat{H}[2] = \sum_{m=1}^{\lfloor M/2 \rfloor} \hat{h}[2m]$ corresponds to the even-odd Trotterization used in one-dimensional tensor network calculations of quantum many-body systems \cite{vidal_efficient_2004}. We consider first-order Trotterization \cite{trotter_1959} of the full imaginary time propagator $e^{-\beta\hat{H}}$:
\begin{align}
e^{-\beta\hat{H}} = \left(\prod_{l=1}^L e^{-\Delta\tau\hat{H}[l]}\right)^{n_\beta} + \mathcal{O}(\Delta\tau^2), \label{eq:e_-betaH}
\end{align}
\noindent where $n_\beta$ is the number of imaginary time steps and $\Delta\tau = \beta/n_\beta$.

The QITE algorithm approximates each imaginary time propagator $e^{-\Delta\tau\hat{H}[l]}$ by a unitary operator
\begin{align}
e^{-i\Delta\tau\hat{G}[l]} = e^{-i\Delta\tau\sum_{\bm{\mu}} x[l]_{\bm{\mu}} \sigma_{\bm{\mu}}},
\end{align}
\noindent where $x[l]_{\bm{\mu}}$ are real coefficients and $\sigma_{\bm{\mu}}$ are Pauli strings. Here we use the notation $\sigma_0 = I, \sigma_x = X, \sigma_y = Y, \sigma_z = Z$ to denote the identity and the Pauli matrices, so that each Pauli string can be written in the form $\sigma_{\bm{\mu}} = \bigotimes_{j=0}^{N-1} \sigma_{\mu_j}$ where $\sigma_{\mu_j}$ acts on qubit $j$ and $\mu_j \in \{0, x, y, z\}$. The Pauli strings  $\sigma_{\bm{\mu}}$ are chosen from the set
\begin{align}
\mathcal{P}_{\hat{H}[l]} = \bigcup_{\hat{h}[m]\in \hat{H}[l]} \mathcal{P}_{\hat{h}[m]},
\label{eq:P_Hl}
\end{align}
where $\mathcal{P}_{\hat{h}[m]}$ is the set all Pauli strings over a domain of $D$ qubits larger than or equal to the support of $\hat{h}[m]$. To apply the QITE unitaries, without an efficient decomposition scheme each unitary needs to be further Trotterized as
\begin{align}
e^{-i\Delta\tau\hat{G}[l]} = \prod_{\bm{\mu}} e^{-i\Delta\tau x[l]_{\bm{\mu}}\sigma_{\bm{\mu}}} + \mathcal{O}(\Delta\tau^2). \label{eq:qite_unitary_trotterized}
\end{align}

The coefficient vector $\bm{x[l]}$ is found by minimizing the square of the difference between the unitarily evolved state $e^{-i\Delta\tau\hat{G}[l]}\ket{\Psi}$ and the imaginary-time-evolved state $c[l]^{-1/2}e^{-\Delta\tau\hat{H}[l]}\ket{\Psi}$, where $c[l] = ||e^{-\Delta\tau\hat{H}[l]}\ket{\Psi}||^2$. This minimization results in a linear system
\begin{align}
\bm{A[l]}\bm{x[l]} = \bm{b[l]}, \label{eq:Ax=b}
\end{align}
\noindent where
\begin{align}
A[l]_{\bm{\mu}\bm{\nu}} &= \Re \langle \Psi | \sigma_{\bm{\mu}} \sigma_{\bm{\nu}}| \Psi \rangle, \label{eq:A} \\
b[l]_{\bm{\mu}} &= \frac{\Im \langle \Psi | e^{-\Delta\tau\hat{H}[l]} \sigma_{\bm{\mu}} | \Psi\rangle}{\Delta\tau c[l]^{1/2}}.
\end{align}
In our implementation, we expand the exponential $e^{-\Delta\tau\hat{H}[l]}$ in $\bm{b[l]}$ and $c[l]$ to second order in $\Delta\tau$:
\begin{align}
b[l]_{\bm{\mu}} &= \frac{\Im \langle \Psi| (- \hat{H}[l] + \Delta\tau \hat{H}[l]^2) \sigma_{\bm{\mu}}|\Psi \rangle}{c[l]^{1/2}} + \mathcal{O}(\Delta\tau^2),\label{eq:b}\\
c[l] &= \langle \Psi | 1 - 2\Delta\tau \hat{H}[l] + {2\Delta\tau^2} \hat{H}^2[l] | \Psi \rangle + \mathcal{O}(\Delta\tau^3) \label{eq:c},
\end{align}

\noindent  To construct the linear systems, given the terms in \cref{eq:A,eq:b,eq:c} we measure operators of the form
\begin{gather}
\sigma_{\bm{\mu}}\sigma_{\bm{\nu}}, \; \hat{H}[l] \sigma_{\bm{\mu}}, \; \hat{H}[l]^2 \sigma_{\bm{\mu}}, \; \hat{H}[l], \; \hat{H}[l]^2.
\label{eq:qite_operators}
\end{gather}

The QITE algorithm is carried out by iterating the procedure of constructing the circuit from the QITE unitaries obtained at the previous imaginary time steps, measuring the operators in \cref{eq:qite_operators}, constructing the linear system in \cref{eq:Ax=b}, solving for $\bm{x[l]}$, and propagating the state with the new unitary $e^{-i\Delta\tau\hat{G}[l]}$.

\subsection{Finite-temperature dynamical\\ correlation functions} 
\label{sec:finite-temperature}

Finite-temperature static observables have been previously computed on quantum hardware using the QMETTS algorithm by averaging over the observable evaluated from each METTS sample state \cite{motta_2019}. In this work, we show that finite-temperature dynamical observables, in particular finite-temperature dynamical correlation functions, can be computed using a similar averaging procedure as for finite-temperature static observables.

\begin{figure}[t]
\centering{
\hspace{-1em}
\includegraphics[width=0.43\textwidth]{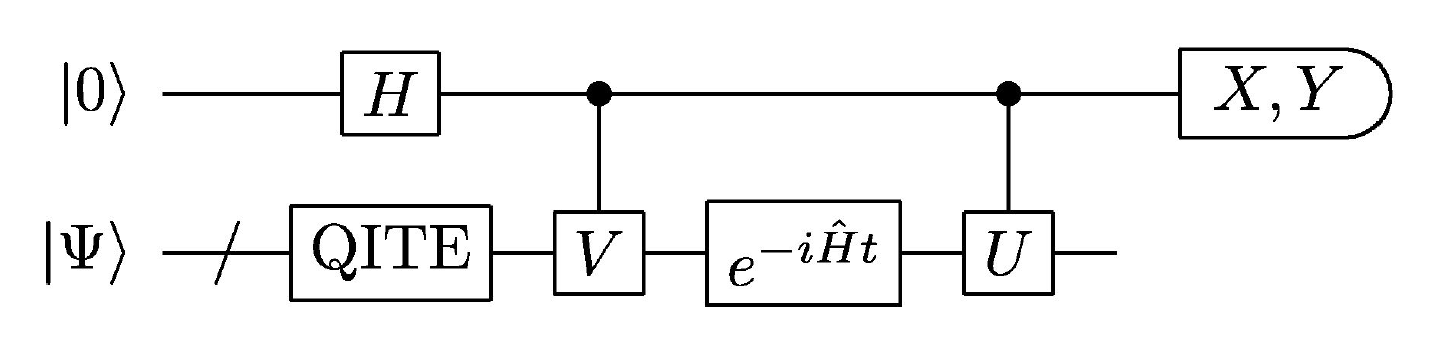}}
\caption{Quantum circuit to calculate the finite-temperature dynamical correlation function $\expval{U(t)V}_\beta$. The ancilla qubit is initialized in $\ket{0}$ and the system qubits are initialized in $\ket{\Psi}$. Measuring $X$ ($Y$) on the ancilla yields the real (imaginary) part of $\expval{U(t)V}$ on the QITE-evolved initial state. Performing a thermal average over all initial states yields $\expval{U(t)V}_\beta$.}
\label{fig:correlation_function_circuit}
\end{figure}

On quantum computers, dynamical correlation functions can be calculated using the circuit reported in Refs.~\cite{ortiz_2001,somma_2002}; this circuit has been recently used to compute neutron scattering cross-section \cite{chiesa_2019} and magnon spectra \cite{francis_2020} on quantum hardware. To obtain finite-temperature dynamical correlation functions, we insert the QITE circuit into the dynamical correlation function circuit, resulting in the circuit shown in \cref{fig:correlation_function_circuit}. The ancilla qubit is initialized in $\ket{0}$ and the system qubits are initialized in $\ket{\Psi}$. Define $\ket{\Psi(\tau)} = e^{-\tau\hat{H}}\ket{\Psi} / || e^{-\tau\hat{H}}\ket{\Psi}||$ as the state initialized in $\ket{\Psi}$ and evolved to imaginary time $\tau$, and $\ket{\Phi(\tau)}$ as the QITE-evolved state that approximates $\ket{\Psi(\tau)}$. Let subscript $a$ ($s$) denote quantities on the ancilla qubit (system qubits). To evaluate finite-temperature observables at an inverse temperature $\beta$, we evolve the initial state by QITE to $\beta / 2$ so that the joint ancilla-system density operator prior to measurement is
\begin{align}
\tilde{\rho} = \tilde{U}(I_a\otimes e^{-i\hat{H}t})\tilde{V}(\rho_a\otimes \rho_{s})\tilde{V}^\dagger(I_a\otimes e^{i\hat{H}t})\tilde{U}^\dagger, \label{eq:rho_tilde}
\end{align}
\noindent where
\begin{gather}
\rho_a = \ket{+}\bra{+},\\
\rho_s = \ket{\Phi(\beta/2)}\bra{\Phi(\beta/2)},
\end{gather}
\noindent and $\tilde{U}$ ($\tilde{V})$ is the controlled-$U$ (controlled-$V$) gate.

Measuring $X$ ($Y$) on the ancilla yields the real (imaginary) part of the dynamical correlation function on a single QITE-evolved basis state:
\begin{align}
\Tr(\tilde{\rho} X_a) &= \Re \langle \Phi(\beta/2) | U(t)V | \Phi(\beta/2)\rangle \\
\Tr(\tilde{\rho} Y_a) &= \Im \langle \Phi(\beta/2) | U(t)V | \Phi(\beta/2)\rangle.
\end{align}

If the initial states are the METTS sample states, an unweighted average over the initial states yields the finite-temperature dynamical correlation function $\expval{U(t)V}_\beta$. In this work, we consider trace evaluation in the exact expression of an observable $\hat{O}$ at finite temperature: 
\begin{align}
\langle \hat{O}\rangle_\beta = \frac{\text{Tr}(e^{-\beta \hat{H}}\hat{O})}{\text{Tr}(e^{-\beta\hat{H}})}. \label{eq:trace}
\end{align}
The numerator trace and the denominator trace are either evaluated by full sampling over the entire Hilbert space, denoted as full trace evaluation, or by random sampling over a subspace of the Hilbert space, denoted as stochastic trace evaluation. If $\hat{O} = U(t)V$, \cref{eq:trace} yields the finite-temperature dynamical correlation function $\expval{U(t)V}_\beta$. $\hat{O}$ can also be a static observable, in which case \cref{eq:trace} yields the static observable at finite temperature.

\section{Methods} 
\label{sec:methods}

\subsection{Pauli string reduction by $\Z_2$ symmetries} \label{sec:pauli_string_reduction}
If we include all $4^D$ Pauli strings over each domain consisting of $D$ qubits, each QITE unitary $e^{-i\Delta\tau\hat{G}[l]}$ applied as in \cref{eq:qite_unitary_trotterized} yields $\mathcal{O}(N4^D)$ multi-qubit rotation gates by the standard rotation gate decomposition \cite{nielsen_and_chuang}, which results in a circuit too deep on near-term quantum devices even for $D = 2$. Because of this prohibitive resource overhead, we describe a systematic method to reduce the number of Pauli strings in the QITE unitaries when the Hamiltonian and initial state have $\Z_2$ symmetries.

$\Z_2$ symmetries on qubit Hamiltonians have direct parallels with the stabilizer formalism in quantum error-correcting codes \cite{gottesman_1998}. Suppose the Hamiltonian has $d$ $\Z_2$ symmetries, i.e. $\hat{H}$ commutes with elements of a group isomorphic to $\Z_2^d$ generated independently by $d$ Pauli strings, and the initial state is in the +1 eigenspace of all $d$ generators. If we regard the symmetry group $\Z_2^d$ as the stabilizer $\mathcal{S}$, the symmetry sector of the initial state corresponds to the stabilizer subspace $V_\mathcal{S}$.

In stabilizer codes, the normalizer of the stabilizer $\mathcal{N}(\mathcal{S})$ includes all Pauli strings that commute with elements of the stabilizer $\mathcal{S}$, and all valid operations on the code space are in the quotient group $\mathcal{N}(\mathcal{S})/\mathcal{S}$. Intuitively, to preserve $\Z_2$ symmetries, among all Pauli strings from $\mathcal{P}_{\hat{H}[l]}$ the QITE unitaries should only include those from the quotient group $\mathcal{N}(\mathcal{S})/\mathcal{S}$. We now show that the original QITE algorithm subsumes the requirement that the Pauli strings should be chosen from $\mathcal{N}(\mathcal{S})/\mathcal{S}$ because the action of the unitary $e^{-i\Delta\tau\hat{G}[l]}$ with the Pauli strings from the unreduced set $\mathcal{P}_{\hat{H}[l]}$ is the same as the action with Pauli strings from the reduced set $\mathcal{P}_{\hat{H}[l]}\cap \mathcal{N}(\mathcal{S})/\mathcal{S}$. This result is stated as the following proposition, proved and discussed in \cref{app:proof_of_pauli_string_reduction}.

\begin{proposition} \label{proposition}
Suppose QITE is applied to approximate the imaginary time propagator $e^{-\Delta\tau\hat{H}[l]}$ on the state $\ket{\Psi}$. If there exists a stabilizer $\mathcal{S}$ such that every element of $\mathcal{S}$ commutes with $\hat{H}[l]$ and $\ket{\Psi}\in V_\mathcal{S}$, then\\
(a) The action of $e^{-i\Delta\tau\hat{G}[l]}$ on $\ket{\Psi}$ with $\sigma_{\bm{\mu}}\in \mathcal{P}_{\hat{H}[l]}$ is equivalent to the action with $\sigma_{\bm{\mu}}\in \mathcal{P}_{\hat{H}[l]}\cap \mathcal{N}(\mathcal{S})/\mathcal{S}$,\\
(b) $e^{-i\Delta\tau\hat{G}[l]} \ket{\Psi} \in V_\mathcal{S}$.
\end{proposition}

Further reduction in the number of Pauli strings can be achieved by recalling from Ref.~\cite{motta_2019} that when the Hamiltonian and the initial state are real in the computational basis, the state after imaginary time evolution must be real. Thus, only Pauli strings with an odd number ofw $Y$ need to be included in the QITE unitaries. Since $\Z_2$ symmetries and the conditions of a real Hamiltonian and initial state are independent, when both conditions are satisfied, the number of Pauli strings can be reduced using both conditions, in which case the reduced set of Pauli strings is modified to $\mathcal{P}_{\hat{H}[l]}\cap \mathcal{N}(\mathcal{S})/\mathcal{S} \cap \{\sigma_{\bm{\mu}}: \sum_j \delta_{\mu_j, y} \equiv 1 \text{ (mod 2)}\}$.

In practice the Proposition is used inductively on the Trotter terms $\hat{H}[l]$, which implies the stabilizer need to be chosen such that every element of the stabilizer commutes with all $\hat{H}[l]$, or equivalently with $\hat{H}$. For spin Hamiltonians, the stabilizer generators are usually global $\Z_2$ symmetries such as $Z^{\otimes N}$ and $X^{\otimes N}$. For general Hamiltonians, the $\Z_2$ symmetries can be found by Gaussian elimination on the parity check matrix formed from the Hamiltonian terms \cite{bravyi_2017}.

\begin{figure}[t]
\centering{
\phantomsubcaption\label{fig:pauli_string_reduction_tfim}
\phantomsubcaption\label{fig:pauli_string_reduction_heisenberg}}
\includegraphics[width=0.48\textwidth]{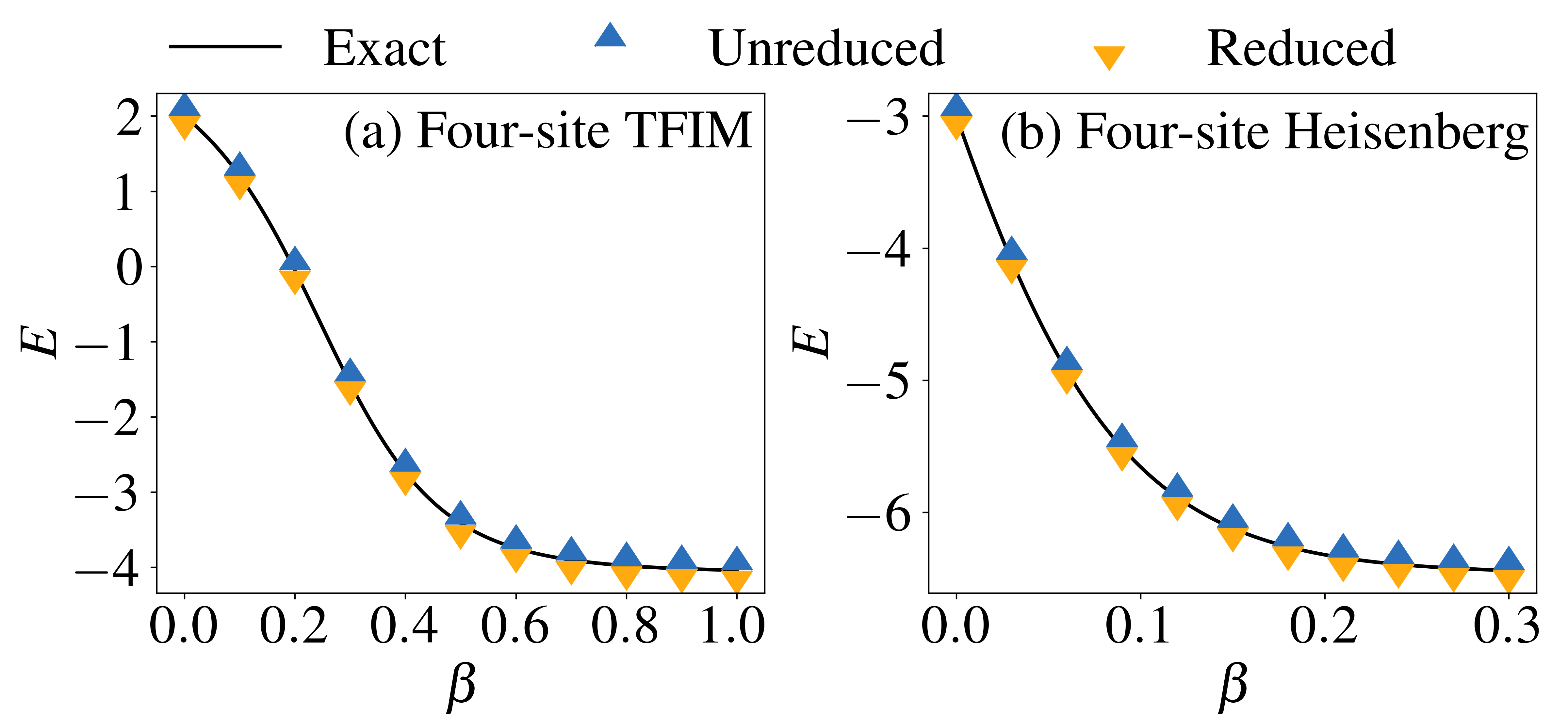}
\caption{Energy $E$ versus imaginary time $\beta$ simulated without noise or measurement sampling on a single initial state with and without reduction of the Pauli strings in the QITE unitaries by $\Z_2$ symmetries. (a) Four-site TFIM with $J = h = 1$ and initial state $\ket{0001}$. The imaginary time step size is set to $\Delta\tau = 0.01$. The number of Pauli strings from three $D = 2$ domains is reduced from 16 to 6 by one $\Z_2$ symmetry $Z_0Z_1Z_2Z_3$.  (b) Four-site Heisenberg model with $J = \Delta = 1$ and initial state $(\ket{0101} + \ket{1010})$ $/\sqrt{2}$. The imaginary time step size is set to $\Delta\tau = 0.03$. The number of Pauli strings on the single $D = 4$ domain is reduced from 120 to 6 by two $\Z_2$ symmetries $Z_0Z_1Z_2Z_3$ and $X_0X_1X_2X_3$. In both panels the energy trajectories using reduced numbers of Pauli strings match the energy trajectories without reduction, which also match the energy trajectories from exact imaginary time evolution.} 
\label{fig:pauli_string_reduction}
\end{figure}

To confirm our Pauli string reduction scheme, we compare the QITE energy trajectory as a function of imaginary time simulated without noise or measurement sampling on a single initial state with and without reduction of the Pauli strings in the QITE unitaries by $\Z_2$ symmetries. The Hamiltonians we study include the transverse-field Ising model (TFIM) Hamiltonian
\begin{align}
\hat{H}_{\text{TFIM}} = J \sum_{i=0}^{N-2} X_i X_{i+1} + h\sum_{i=0}^{N-1} Z_i \label{eq:tfim}
\end{align}
\noindent and the Heisenberg $XXZ$ Hamiltonian
\begin{align}
\hat{H}_{XXZ} = J \sum_{i=0}^{N-2} (X_i X_{i+1} + Y_i Y_{i+1} + \Delta Z_i Z_{i+1}), \label{eq:heisenberg}
\end{align}
\noindent with open boundary conditions assumed for both. 

In \cref{fig:pauli_string_reduction} we plot energy versus imaginary time calculated with QITE on a single initial state. The unreduced set of Pauli strings only includes Pauli strings with odd numbers of $Y$ because the Hamiltonian and initial state are real in the computational basis. We choose a sufficiently small imaginary time step size $\Delta\tau$ to ensure that the Trotter errors from expansion in \cref{eq:qite_unitary_trotterized} are negligible. Figure \ref{fig:pauli_string_reduction_tfim} plots the energy trajectory for the initial state $\ket{0001}$ in the four-site TFIM with $J = h = 1$. The Hamiltonian and the initial state have a $\Z_2$ symmetry $Z_0Z_1Z_2Z_3$. By combining reduced Pauli strings from all three $D = 2$ domains, we obtain six Pauli strings in the QITE unitaries
\begin{align}
X_0Y_1, Y_0X_1, X_1Y_2, Y_1X_2, X_2Y_3, Y_2X_3, \label{eq:6_D2_pauli_strings}
\end{align}
\noindent compared to 16 Pauli strings without reduction by $\Z_2$ symmetries. Figure \ref{fig:pauli_string_reduction_heisenberg} plots energy versus imaginary time of the initial state $(\ket{0101} + \ket{1010})/\sqrt{2}$ on the four-site Heisenberg model with $J = \Delta = 1$. The Hamiltonian and the initial state have two $\Z_2$ symmetries $Z_0Z_1Z_2Z_3$ and $X_0X_1X_2X_3$. The 120 Pauli strings in the QITE unitaries without reduction is reduced to the 6 Pauli strings
\begin{align}
X_0Y_1Z_2, X_0Z_1Y_2, Y_0X_1Z_2, Y_0Z_1X_2, Z_0X_1Y_2, Z_0Y_1X_2. \label{eq:6_D4_pauli_strings}
\end{align}
\noindent In both panels of \cref{fig:pauli_string_reduction}, the energy trajectories using reduced numbers of Pauli strings match the energy trajectories without reduction, which also match the energy trajectories from exact imaginary time evolution.

\subsection{Circuit optimization}
\label{sec:circuit_optimization}

Even with reduction of Pauli strings in the QITE unitaries by $\Z_2$ symmetries, applying the QITE unitaries as in \cref{eq:qite_unitary_trotterized} may still result in a circuit too deep to be implemented on current quantum hardware. In this section we describe circuit optimization techniques that further reduce circuit depth.

In two-site calculations, both the QITE circuit and the real time evolution circuit can be optimized to constant depth with a standard one- and two-qubit gate set, regardless of the number of imaginary and real time steps. For example, in two-site TFIM there is only one Pauli string $X_0Y_1$ in the QITE unitaries after reduction by the $\Z_2$ symmetry $Z_0Z_1$. Suppose the unitary applied to the state at the $k$-th imaginary time step is $e^{-i\Delta\tau x_k X_0Y_1}$. Then the unitaries at all imaginary time steps can be multiplied into a single two-qubit rotation gate $e^{-i \theta X_0Y_1 / 2}$ where $\theta = 2 \Delta\tau \sum_k x_k$. For real time evolution, the two-qubit operator $e^{-i\hat{H}t}$ is decomposed by the $KAK$ decomposition \cite{khaneja_2001,kraus_2001,vatan_2004,vidal_universal_2004} into six single-qubit gates and two CNOT gates.

\begin{figure}[t]
\centering{
\includegraphics[width=0.33\textwidth]{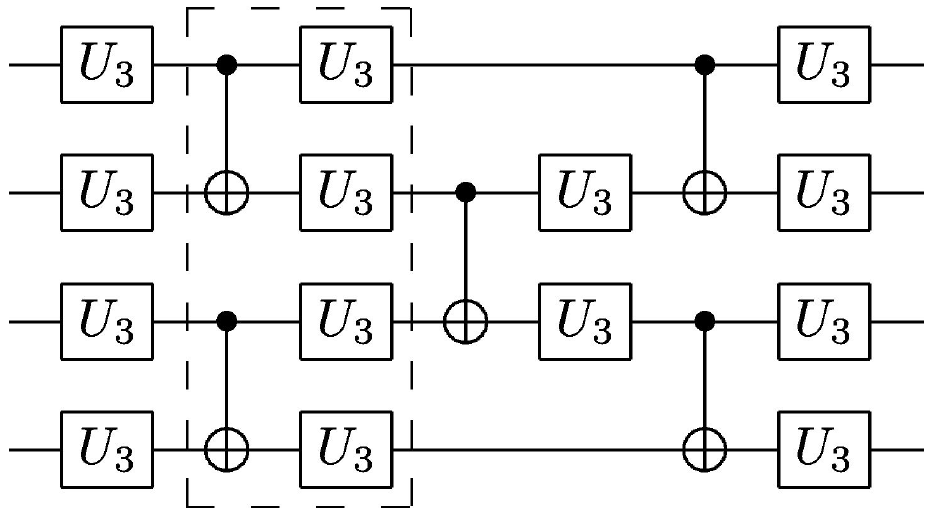}}
\caption{Four-site recompiled circuit. The four $U_3$ gates at the left constitute the base gate round. Each additional gate round includes a layer of CNOT gates and a layer of single-qubit gates as shown in the dashed box. The additional gate rounds alternate between even-odd and odd-even pairs of qubits, so that the circuit shown consists of three gate rounds.}
\label{fig:recompilation}
\end{figure}

In four-site calculations, neither the QITE circuit nor the real time evolution circuit is of constant depth. If we Trotterize the QITE unitaries as in \cref{eq:qite_unitary_trotterized} and similarly for the real time propagator, the circuit is too deep to be accurately implemented on existing quantum devices. Therefore, we recompile the circuit by fitting each QITE unitary $e^{-i\Delta\tau\hat{G}[l]}$ or the real time propagator $e^{-i\hat{H}t}$ to a parametrized circuit \cite{khatri_2019,jones_2018,heya_2018} using an open-source tensor network-based quantum simulation library \cite{quimb}. In \cref{fig:recompilation}, we show the recompiled four-site QITE circuit, where the $U_3$ gate is a generic single-qubit gate defined as
\begin{align}
U_3(\theta, \phi, \lambda) = 
\begin{pmatrix}
\cos(\theta/2) & -e^{i\lambda} \sin(\theta/2) \\
e^{i\phi}\sin(\theta/2) & e^{i(\lambda + \phi)}\cos(\theta/2)
\end{pmatrix}.
\end{align}
\noindent The four $U_3$ gates at the left constitute the base gate round. Each additional gate round consists of a layer of CNOT gates and a layer of single-qubit gates. The additional gate rounds alternate between even-odd and odd-even pairs of qubits. Let the target unitary be $\mathcal{U}_{\text{targ}}$ and the recompiled unitary be $\mathcal{U}_{\text{rec}}(\bm{\theta})$, where $\bm{\theta}$ is a composite variable denoting all the angles. Given a reduced density operator $\rho$ on the finite domain acted on by the target unitary, write $\rho_{\text{targ}} = \mathcal{U}_{\text{targ}}\rho\,\mathcal{U}_{\text{targ}}^\dagger$ and $\rho_{\text{rec}}(\bm{\theta}) = \mathcal{U}_{\text{rec}}(\bm{\theta})\rho\,\mathcal{U}_{\text{rec}}^\dagger(\bm{\theta})$. The optimal recompiled unitary is found by performing a gradient descent to maximize the fidelity \cite{jozsa_1994}
\begin{align}
F(\bm{\theta}) = \left(\text{Tr} \sqrt{\rho_{\text{targ}}^{1/2}\rho_{\text{rec}}(\bm{\theta})\rho_{\text{targ}}^{1/2}}\right)^2.
\label{eq:fidelity}
\end{align}

Since the QITE unitaries are real, we use the one-parameter single-qubit gate $R_y(\theta) = U_3(\theta, 0, 0)$ in the recompiled circuit for QITE, while for real time evolution we keep the $U_3$ gate as the parametrized single-qubit gate.

\subsection{Error mitigation}
\label{sec:error_mitigation}

To mitigate the effect of hardware noise on the measurement results, we post-process our hardware data by error mitigation methods including post-selection, readout error mitigation and phase-and-scale correction. Post-selection and readout error mitigation are applied to the measurement outcomes at each imaginary time step in the QITE subroutine; phase-and-scale correction is applied to the final computed finite-temperature dynamical correlation function as a single-step post-processing.

Post-selection is performed on $\Z_2$ symmetries discussed in \cref{sec:pauli_string_reduction}. When the Hamiltonian and the initial state have $\Z_2$ symmetries, the final state after imaginary or real time evolution should have the same stabilizer parities as the initial state. However, during execution of the circuit, gate errors and qubit decoherence can induce nonzero overlap of the qubit state with the subspace of the wrong parity. Post-selection can mitigate these undesirable effects by discarding measurement outcomes with the wrong parity \cite{bonet-monroig_2018,mcardle_error-mitigated_2019}.

\begin{figure}[b]
\centering{
\includegraphics[width=0.37\textwidth]{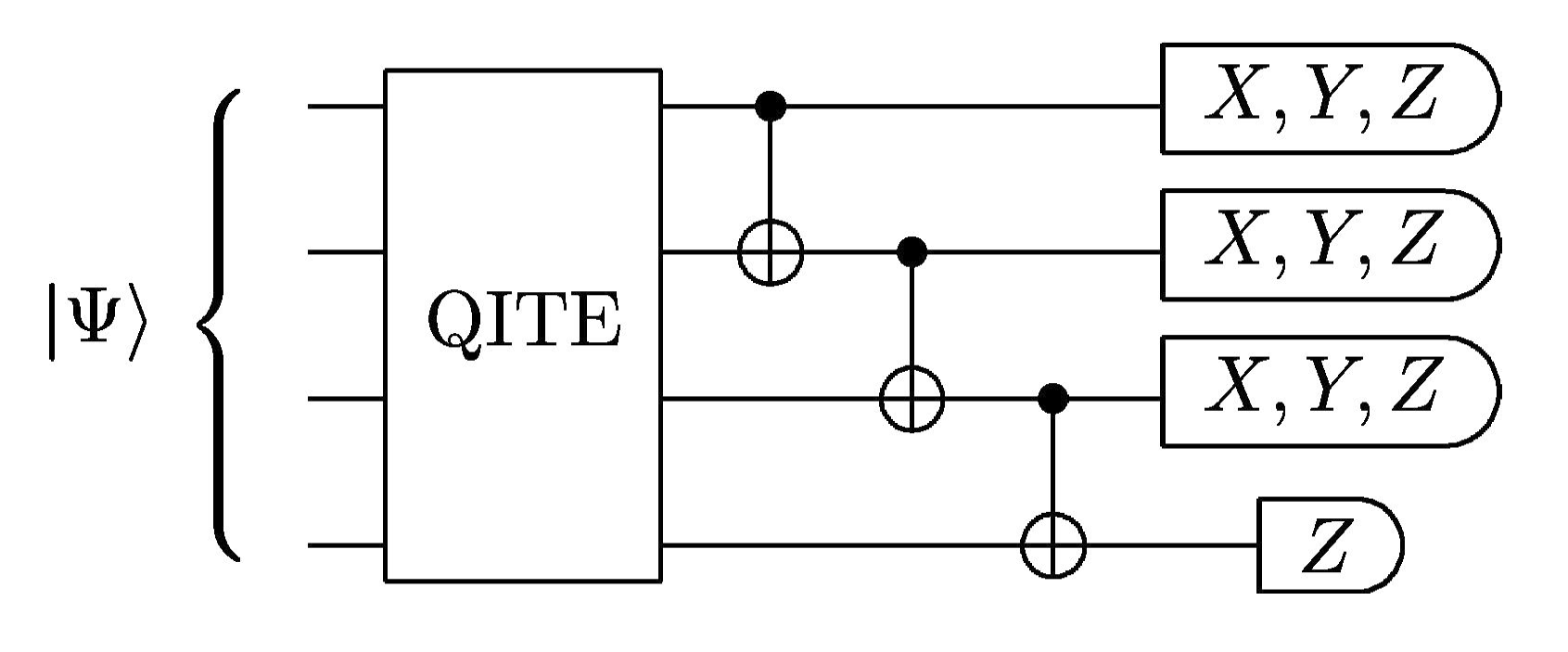}}
\caption{Measurement of a Pauli string in a four-site QITE calculation with post-selection on the stabilizer generator $Z_0Z_1Z_2Z_3$. The appended CNOT gates achieve simultaneous measurement of the Pauli string with the stabilizer generator by transforming $Z_0Z_1Z_2Z_3$ to $Z_3$ acting on a single qubit, from which the stabilizer parity is read off. The other qubits are measured in $X$-, $Y$- or $Z$-basis depending on the Pauli string measured. Measurement outcomes with the wrong parity are discarded.}
\label{fig:post_selection}
\end{figure}

We specifically consider the symmetry from a single stabilizer generator.  If the operator to be measured is an ancilla operator, we can simply measure the stabilizer generator on all the system qubits and read off the parity without interfering with measurement of the ancilla. If the operator to be measured acts on system qubits, we need to simultaneously measure the operator and the stabilizer generator, which is possible because all operators in \cref{eq:qite_operators} commute with the stabilizer generator by our choice of Pauli strings in the QITE unitaries in \cref{sec:pauli_string_reduction}. Specifically, each operator and the stabilizer generator can be simultaneously measured by using Clifford gates to transform the Pauli string components of the operator and the stabilizer generator until they are qubit-wise commuting, so that their expectation values can be read off on different qubits \cite{gokhale_2019,crawford_2019,yen_2020,hamamura_2020}. 

Figure \ref{fig:post_selection} shows the circuit to simultaneously measure the stabilizer generator $Z_0Z_1Z_2Z_3$ and a Pauli string that commutes with it in a four-site QITE calculation. The sequence of CNOT gates after the QITE circuit in \cref{fig:post_selection} transforms $Z_0Z_1Z_2Z_3$ to $Z_3$. Since $Z_3$ acts on a single qubit, it necessarily qubit-wise commutes with the transformed Pauli string. In practice, we stop applying CNOT gates when the transformed Pauli string becomes qubit-wise commuting with the transformed stabilizer generator, and therefore the number of CNOT gates added to the end of the circuit ranges from zero to three.

To account for noise in the final measurement, the built-in readout error mitigation routine in Qiskit \cite{qiskit_textbook} is applied to each measurement outcome. Because of the small size of the systems we study, for $N$ qubits we carry out full calibration on all $2^N$ initial states. Application of the inverse of the calibration matrix to the raw measurement counts is performed by the default least-square fitting method.

\begin{figure}[t]
\centering{
\phantomsubcaption\label{fig:error_mitigation_2-site}
\phantomsubcaption\label{fig:error_mitigation_4-site}
\includegraphics[width=0.48\textwidth]{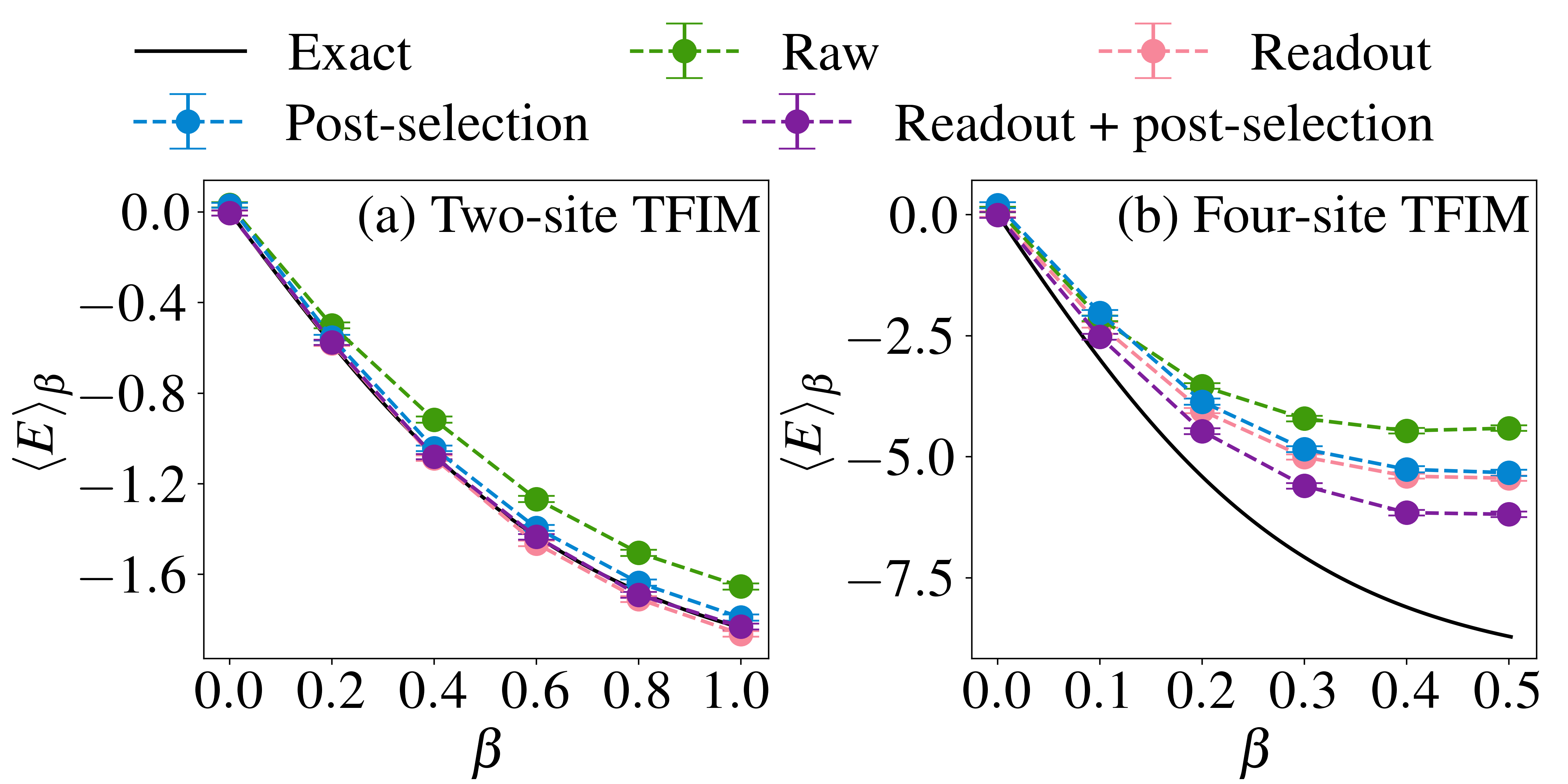}}
\caption{Finite-temperature energy $\expval{E}_\beta$ of (a) the two-site TFIM with $J = h = 1$ and (b) the four-site TFIM with $J = 3$, $h = 1$, simulated with measurement sampling and the noise model from \textit{ibmq\_rome}. The imaginary time step size is set to $\Delta\tau = 0.1$. Raw data are post-processed at each imaginary time step with either readout error mitigation, or post-selection, or both. Employing both readout error mitigation and post-selection is observed to be most effective in mitigating the errors.}
\label{fig:error_mitigation}
\end{figure}

We assess the effectiveness of applying post-selection and readout error mitigation at every imaginary step of QITE by simulating the finite-temperature energies of two-site and four-site TFIMs using full trace evaluation with measurement sampling and the noise model from \textit{ibmq\_rome}. In both panels of \cref{fig:error_mitigation}, QITE is applied with Pauli strings reduced and circuits optimized. In particular, four-site QITE unitaries are of domain size $D = 2$ and recompiled with three rounds of gates. From \cref{fig:error_mitigation} we can see that both readout error mitigation and post-selection shift the raw data toward the exact data, confirming the effectiveness of both schemes in reducing the effect of noise. Furthermore, a combination of readout error mitigation and post-selection is observed to be most effective in mitigating the errors, which is not apparent on two sites in \cref{fig:error_mitigation_2-site} presumably because of the small size of the system but clearly evident on four sites in \cref{fig:error_mitigation_4-site}.

In calculations of finite-temperature dynamical correlation functions, the ancilla qubit is in the state $\ket{+}$ before entangling with the system qubits. When there is a long sequence of gates in the real time propagator $e^{-i\hat{H}t}$, decoherence of the ancilla qubit such as amplitude damping to the qubit ground state $\ket{0}$ and depolarization will significantly affect the $X$ and $Y$ measurement results on the ancilla. To mitigate the effect of ancilla decoherence, we apply phase-and-scale correction \cite{chiesa_2019,francis_2020} as a single-step post-processing to the result at the end of the calculation. The only finite-temperature dynamical correlation function considered in this work is $\expval{Z_0(t)Z_0}_\beta$, which is equal to 1 analytically at $t = 0$. Hence, we apply phase-and-scale correction by dividing the raw hardware $\expval{Z_0(t)Z_0}_\beta$ at each $t$ by the raw hardware $\expval{Z_0(t=0)Z_0}_\beta$ to enforce the condition $\expval{Z_0(t=0)Z_0}_\beta = 1$.

\section{Results}
\label{sec:results}

Experiments of computing finite-temperature observables were conducted on IBM Quantum devices \textit{ibmq\_bogota} and \textit{ibmq\_rome} \cite{ibmq_5-qubit}, both of which consist of five qubits arranged on a chain with nearest-neighbor interactions and similar error rates. IBM's open-source library Qiskit \cite{qiskit} was used to implement our algorithms on the devices. In each calculation, the $N$ system qubits $0, ..., N-1$ are arranged adjacent to each other and the ancilla is closest to system qubit 0. 

The systems we study are sufficiently small that we apply QITE to approximate the full imaginary time propagator $e^{-\Delta\tau\hat{H}}$ at each imaginary time step, which is equivalent to setting $L = 1$ and $\hat{H}[1] = \hat{H}$ in \cref{eq:e_-betaH}. The QITE linear systems in \cref{eq:Ax=b} are solved by a conjugate gradient method. Because hardware noise and measurement sampling lead to ill-conditioned $\bm{A}$ matrices in the QITE linear systems, we add a regularizer of 0.2 to the diagonal elements of each $\bm{A}$ matrix in the four-site calculations.

Each calibration circuit used for readout error mitigation is repeated 1000 times; each Pauli string measurement circuit used to construct the QITE linear systems is repeated 8000 times. Error bars from full trace evaluation result only from measurement sampling and are the size of the markers in most figures; error bars from stochastic trace evaluation originate from both measurement sampling and initial state sampling. A detailed description of error bars in full and stochastic trace evaluation is given in \cref{app:error_bars}.

\subsection{Two-site calculations}
\label{sec:2-site}

\begin{figure}[t]
\centering{
\phantomsubcaption\label{fig:2-site_energy}
\phantomsubcaption\label{fig:2-site_x0x1}
\includegraphics[width=0.48\textwidth]{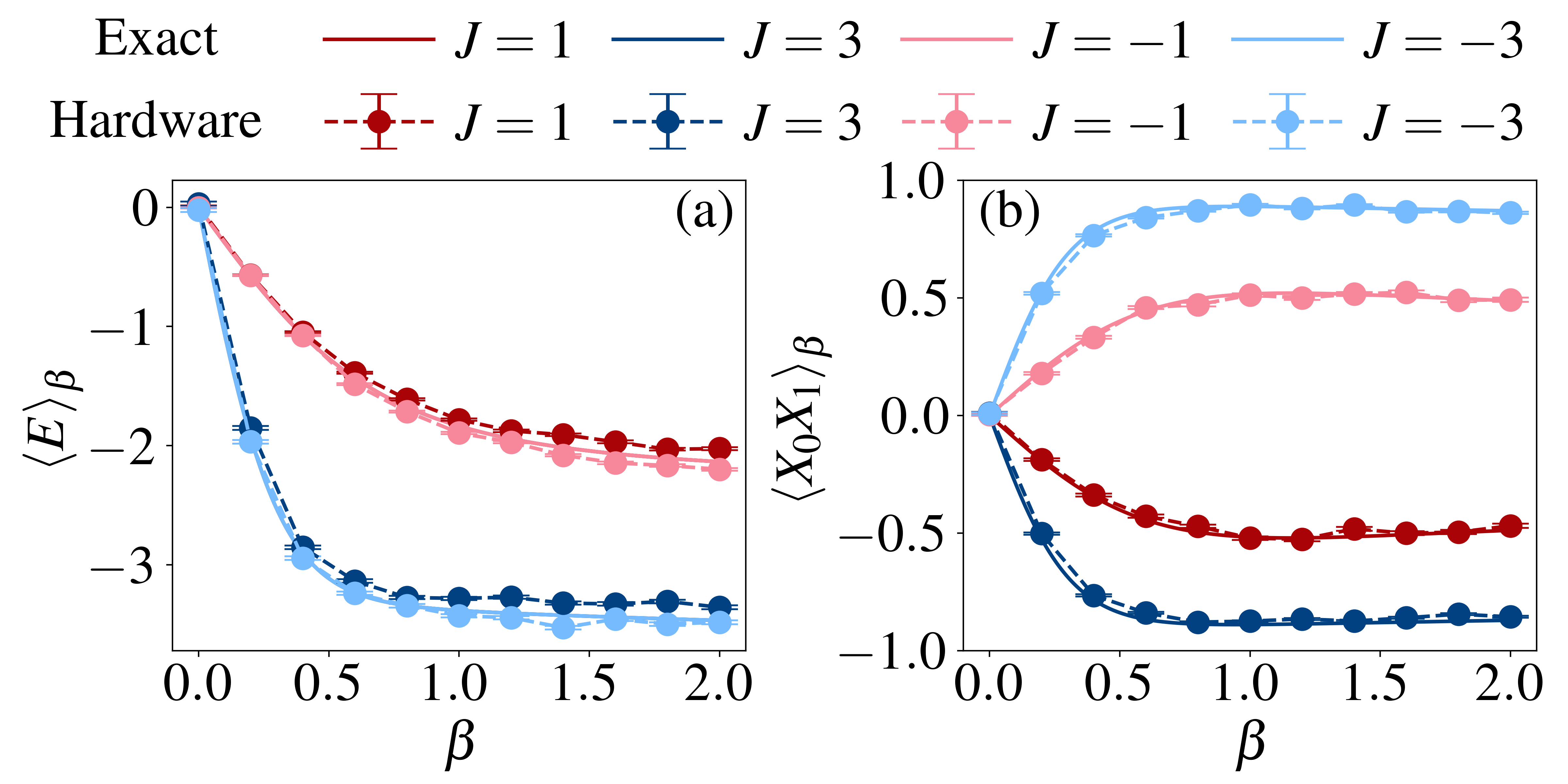}}
\caption{(a) Finite-temperature energy  $\expval{E}_\beta$ and (b) static correlation function $\expval{X_0X_1}_\beta$ of the two-site TFIM with $J = \pm 1, \pm 3$ and $h = 1$ versus inverse temperature $\beta$. The imaginary time step size is set to $\Delta\tau = 0.1$. In each of the calculated observables, the mean absolute percentage error between hardware and exact results averaged over all $\beta$ ranges from 1\% to 4\%.}
\label{fig:2-site_static}
\end{figure}

We study the two-site TFIM defined in \cref{eq:tfim} by setting $h = 1$ and varying $J$. The finite-temperature energy $\expval{E}_\beta$ and static correlation function $\expval{X_0X_1}_\beta$ are calculated on the Hamiltonians with $J = \pm 1, \pm 3$, while the finite-temperature dynamical correlation function $\expval{Z_0(t)Z_0}_\beta$ and excitation spectra are calculated on the Hamiltonian with $J = 3$. In all calculations, finite-temperature observables are calculated by full trace evaluation and the circuits are optimized according to the circuit optimization procedures in \cref{sec:circuit_optimization}.

Figure \ref{fig:2-site_static} shows the finite-temperature energy $\expval{E}_\beta$ and static correlation function $\expval{X_0X_1}_\beta$ of the two-site TFIM with $J = \pm 1, \pm 3$ from $\beta = 0$ to $\beta = 2$. In both \cref{fig:2-site_energy} and \cref{fig:2-site_x0x1} the finite-temperature observables obtained on hardware are in good agreement with exact values. Further, if we regard each finite-temperature variable as a function of $J$, analytically it can be shown that $\expval{E}_{\beta}(J) = \expval{E}_{\beta}(-J)$ and $\expval{X_0X_1}_{\beta}(J) = - \expval{X_0X_1}_{\beta}(-J)$. This relation is satisfied in the hardware data. In each of the observables we calculated, the mean absolute percentage error between hardware and exact results averaged over all $\beta$ ranges from 1\% to 4\%.

\begin{figure*}[h]
\centering
{\phantomsubcaption\label{fig:2-site_real}
\phantomsubcaption\label{fig:2-site_imag}
\phantomsubcaption\label{fig:2-site_spec}
\phantomsubcaption\label{fig:2-site_peaks}
\includegraphics[width=0.92\textwidth]{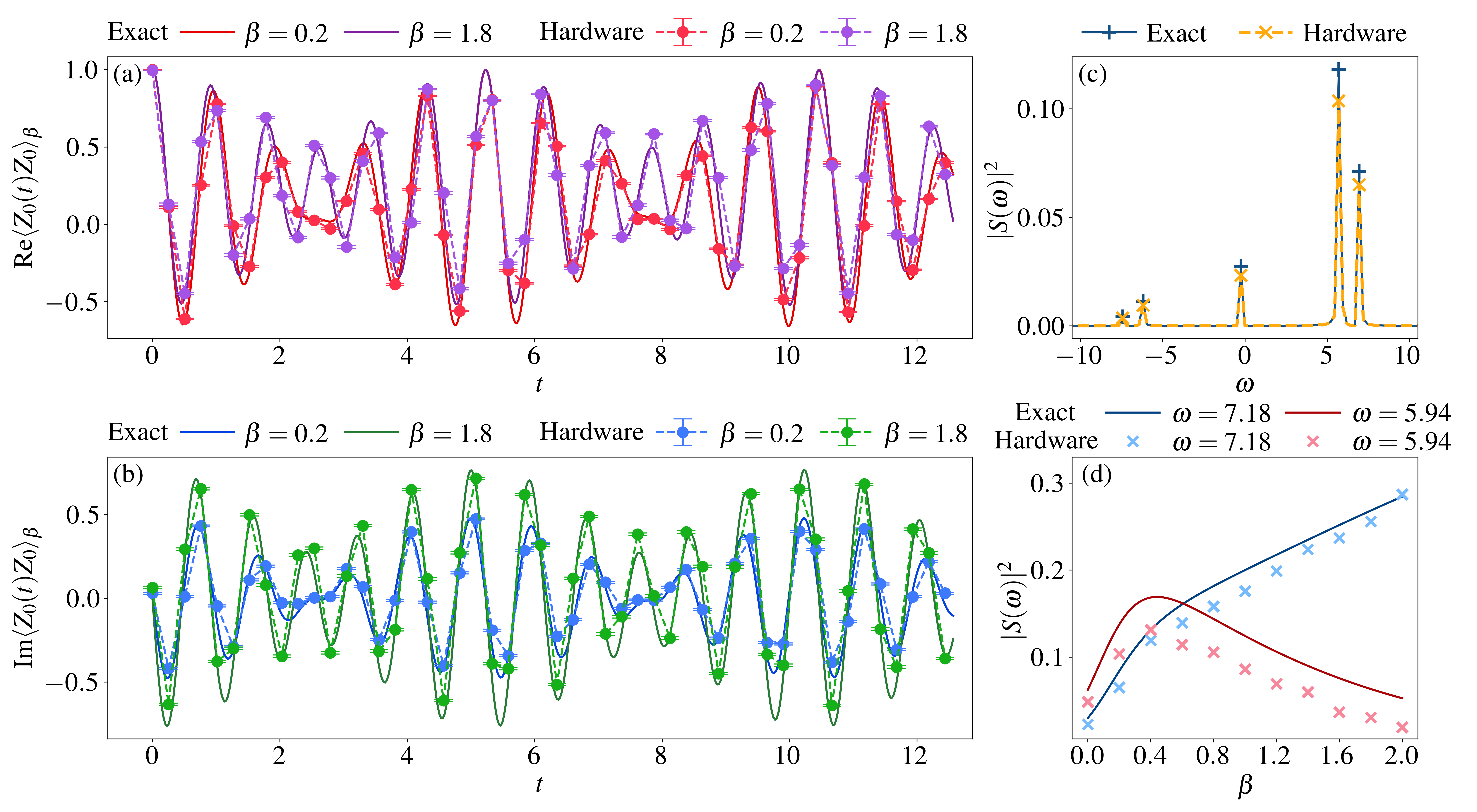}}
\caption{Finite-temperature dynamical properties of the two-site TFIM with $J = 3$, $h = 1$. The imaginary time step size is set to $\Delta\tau = 0.1$. (a) Real and (b) imaginary parts of the finite-temperature dynamical correlation function $\expval{Z_0(t)Z_0}_\beta$ at $\beta = 0.2$ and $\beta = 1.8$ versus real time $t$. (c) Finite-temperature excitation spectra $|S(\omega)|^2$ versus frequency $\omega$. Positive (negative) frequencies correspond to emissions (absorptions). (d) Amplitudes of the two emission peaks at $\omega = 7.18$ and $\omega = 5.94$. The hardware data capture finite-temperature dynamics of two-site TFIM across a wide range of temperatures.}
\label{fig:2-site_dynamical}
\end{figure*}

Next, finite-temperature dynamical properties were calculated on the two-site TFIM with $J = 3, h = 1$. The dynamical correlation function $\expval{Z_0(t)Z_0}_\beta$ is evaluated from $\beta = 0$ to $\beta = 2$ and at real time from $t = 0$ to $t = 8\pi$. Figures \ref{fig:2-site_real} and \ref{fig:2-site_imag} show the real and imaginary parts of $\expval{Z_0(t)Z_0}_\beta$ at $\beta = 0.2$ and $\beta = 1.8$ up to $t = 4\pi$. From \cref{fig:2-site_real,fig:2-site_imag} we see that even without phase-and-scale correction, the real and imaginary parts of $\expval{Z_0(t)Z_0}_\beta$ agree well with the exact results at both small and large $\beta$, presumably due to the constant and shallow depth of the real time evolution circuit.

The spectral density $S(\omega)$ is obtained by a discrete Fourier transform of the dynamical correlation function 
\noindent $\expval{Z_0(t)Z_0}_\beta$. Specifically, at each $\beta$
\begin{align}
S(\omega_k) = \frac{1}{n_t} \sum_{m=0}^{n_t-1} \expval{Z_0(t_m)Z_0}_\beta e^{i\omega_k t_m}, \label{eq:fft}
\end{align}
\noindent where $n_t$ is the total number of points in the time series, $t_m = m\Delta t$, and $\omega_k = 2\pi k / n_t \Delta t$. With this definition of Fourier transform, the peaks at positive (negative) frequencies correspond to emissions (absorptions) of excitations of the system.

In \cref{fig:2-site_spec} we plot the excitation spectra of the two-site TFIM at $\beta = 0.2$. The exact excitation spectrum is obtained by a Fourier transform of the exact $\expval{Z_0(t)Z_0}_\beta$ at the same points in real time as the $\expval{Z_0(t)Z_0}_\beta$ obtained on hardware. From the plot, we can see that the hardware excitation spectrum agrees well with the exact

\onecolumngrid
\vspace{1.0em}

\twocolumngrid

\noindent excitation spectrum. The frequencies $\omega = 0$, $\pm 5.94$, $\pm 7.18$, at which the peaks in the excitation spectra are located, correspond to the excitation frequencies of $\omega = 0, \pm 6.00, \pm 7.21$ from exact diagonalization of the Hamiltonian. The deviation of the frequencies in the excitation spectra obtained from hardware $\expval{Z_0(t)Z_0}_\beta$ compared to the frequencies obtained from exact diagonalization is due to the finite real time domain in our hardware calculations.

To analyze the evolution of the excitation spectra across different temperatures, we plot the amplitudes at the two emission frequencies versus $\beta$ in \cref{fig:2-site_peaks}. Analytically, the amplitude of the transition from an initial state $\ket{\Psi_i}$ to a final state $\ket{\Psi_f}$ is $e^{-\beta E_f}|\langle \Psi_i| Z_0 | \Psi_f\rangle |^2 / \mathcal{Z}$, where $E_f$ is the energy of the final state and $\mathcal{Z}$ is the partition function. In the two-site TFIM, the only allowed transitions are between the two states in each of the two-dimensional eigenspaces of $Z_0Z_1$ with eigenvalues $\pm 1$. The frequency $\pm 7.18$ corresponds to a transition in the $+1$ eigenspace, where the ground state lies, and the frequency $\pm 5.94$ corresponds to a transition in the $-1$ eigenspace, where the first excited state lies. As the temperature decreases from infinite temperature ($\beta$ increases from 0), the populations in the two lowest states first increase until the ground state population dominates over that of the first excited state at around $\beta = 0.4$, a trend reproduced by the amplitudes obtained from hardware data in \cref{fig:2-site_peaks}. Thus, \cref{fig:2-site_dynamical} shows that quantum hardware accurately captures the finite-temperature dynamics of the two-site TFIM across a wide range of temperatures.

\subsection{Four-site calculations}
\label{sec:4-site}

We next proceed to four-site spin systems. We study the four-site TFIM defined in \cref{eq:tfim} with $J = 3, h = 1$. Full trace evaluation is employed unless otherwise specified.

First, let us consider the gate count in the four-site circuits. For the four-site TFIM with $D = 2$, after reduction by the $\Z_2$ symmetry $Z_0Z_1Z_2Z_3$ there are six weight-two Pauli strings, which are the ones given in \cref{eq:6_D2_pauli_strings}. If we Trotterize the QITE unitaries as in \cref{eq:qite_unitary_trotterized}, each unitary requires 12 CNOT gates by the standard rotation gate decomposition \cite{nielsen_and_chuang}, which becomes unfeasible on near-term quantum hardware after the first few imaginary time steps. When $D = 4$, even after reduction by one $\Z_2$ symmetry there are still 28 Pauli strings in each QITE unitary. After Trotterization and rotation gate decomposition, each QITE unitary requires more than 50 CNOT gates for a single imaginary time step. Hence, when $D = 2$, we compute finite-temperature observables both by Trotterizing and by recompiling the QITE unitaries with three gate rounds; when $D = 4$, we only recompile the QITE unitaries with three gate rounds. To obtain dynamical correlation functions, we additionally recompile the real time propagator $e^{-i\hat{H}t}$ with five gate rounds. The number of gate rounds is chosen so that the fidelity in \cref{eq:fidelity} is at least 0.999 on average in each calculation.  

\begin{figure}[b]
\centering
\includegraphics[width=0.48\textwidth]{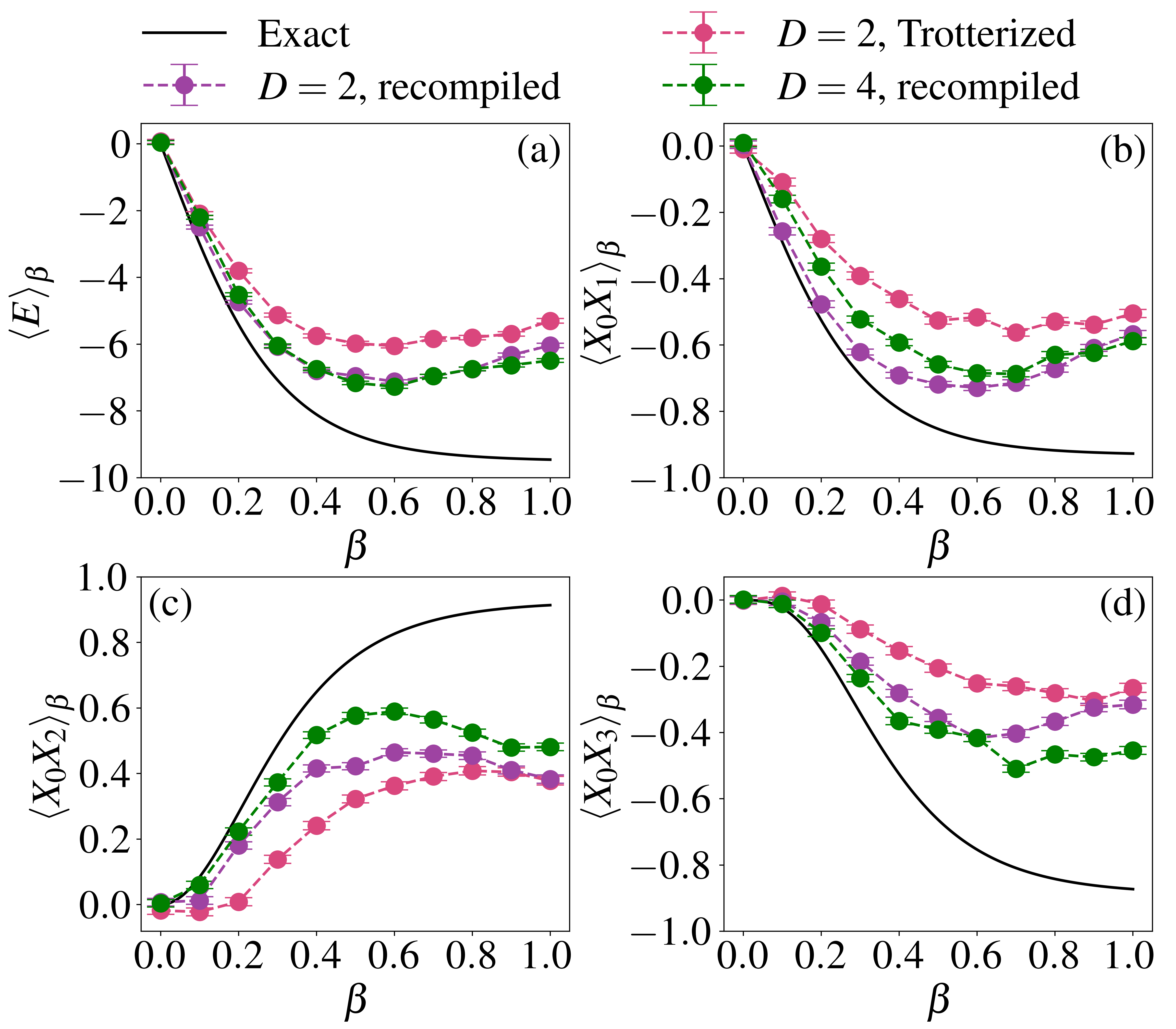}
\caption{(a) Finite-temperature energy $\expval{E}_\beta$ and static correlation functions (b) $\expval{X_0X_1}_\beta$ (c) $\expval{X_0X_2}_\beta$ (d) $\expval{X_0X_3}_\beta$ of the four-site TFIM with $J = 3, h = 1$ versus inverse temperature $\beta$ with different QITE unitaries. The imaginary time step size is set to $\Delta\tau = 0.05$. The $D = 2$ QITE unitaries are either Trotterized as in \cref{eq:qite_unitary_trotterized} or recompiled, while all $D = 4$ QITE unitaries are recompiled. The results with recompiled QITE unitaries are closer to exact results than the results with Trotterized QITE unitaries due to circuit depth. Between the calculations with recompiled unitaries, $D = 4$ is not necessarily closer to exact results than $D = 2$ for all observables possibly due to the increased influence of hardware noise in the larger linear systems.}
\label{fig:4-site_static}
\end{figure}

Figure \ref{fig:4-site_static} shows the finite-temperature energy $\expval{E}_\beta$ and static correlation functions $\expval{X_0X_1}_\beta, \expval{X_0X_2}_\beta, \expval{X_0X_3}_\beta$ of the four-site TFIM. From the figure, we can see that the finite-temperature observables calculated with Trotterized $D = 2$ QITE unitaries deviate from those calculated with $D = 2$ recompiled QITE unitaries or $D = 4$ recompiled QITE unitaries after $\beta = 0.1$. This deviation is due to the deep circuit resulting from 12 layers of CNOT gates per imaginary time step, compared to 3 layers of CNOT gates per imaginary time step in the recompiled circuit. Moreover, even the recompiled QITE unitaries are not able to track the exact finite-temperature observables after the first few $\beta$. In particular, the slope is reversed compared to the exact result after $\beta = 0.5$. QITE up to $\beta = 0.5$ corresponds to 5 imaginary time steps and hence 15 layers of CNOT gates, which is almost at the limit of circuit depth on these quantum devices. 

We examine more closely the calculations with recompiled QITE unitaries and focus on the data at $\beta \leq 0.5$. For both $D = 2$ and $D = 4$, the recompiled QITE unitaries apply the same number of layers of gates at each imaginary time step, so the difference in calculated observables is formally caused by the difference in domain sizes. Since the domain size should grow with correlation length and hence with imaginary time \cite{motta_2019}, we should expect the hardware results to be closer to the exact results with $D = 4$ than with $D = 2$ especially at large $\beta$. However, this hypothesis only holds for $\expval{X_0X_2}_\beta$ and $\expval{X_0X_3}_\beta$ but not for $\expval{E}_\beta$ and $\expval{X_0X_1}_\beta$. Failure of this hypothesis is likely due to that fact that the larger 28-dimensional linear system arising from $D = 4$ incorporates more errors from hardware noise compared to the smaller 6-dimensional linear system arising from $D = 2$.

\begin{figure}[t]
\centering
\includegraphics[width=0.44\textwidth]{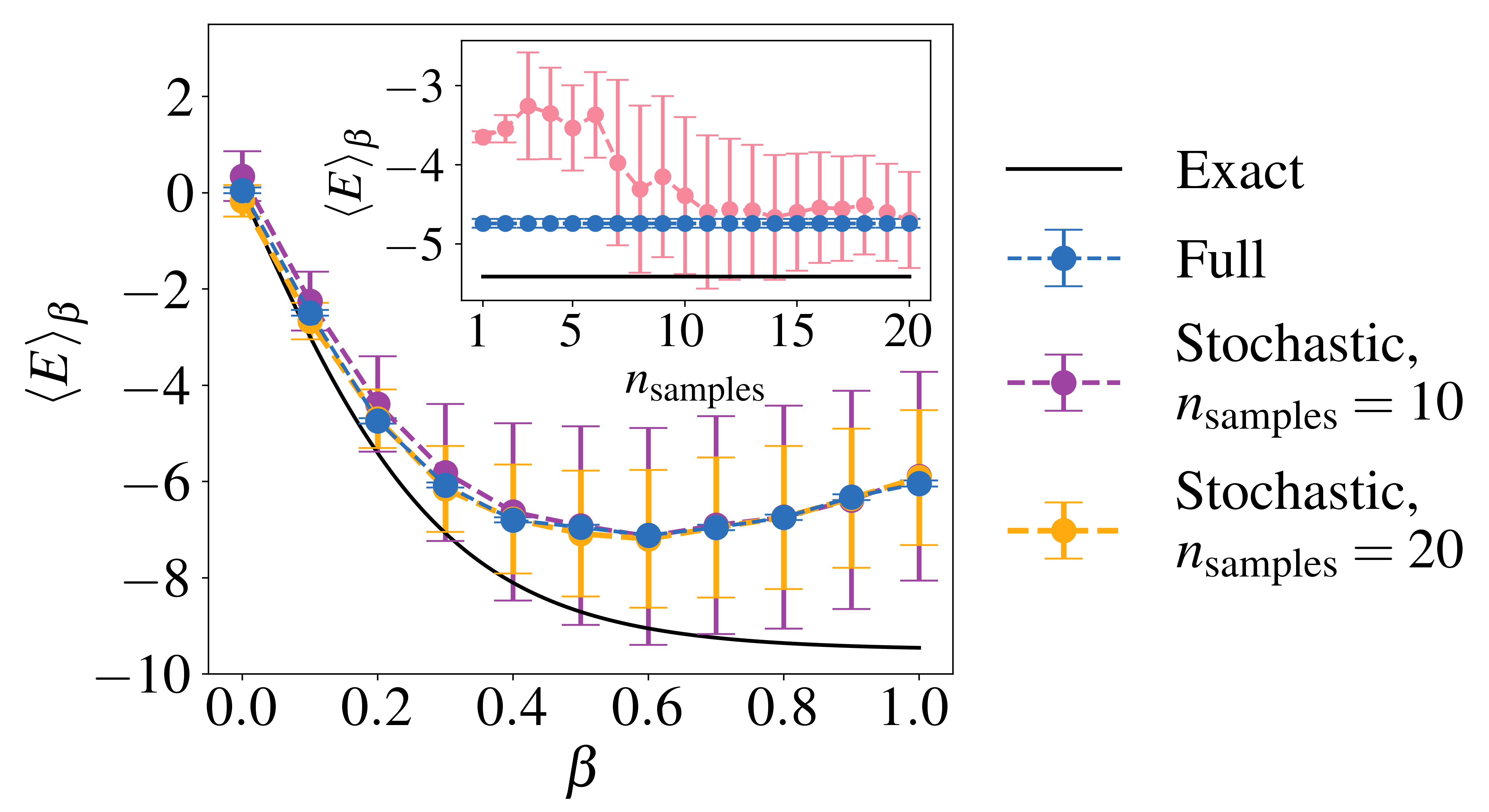}
\caption{Finite-temperature energy $\expval{E}_\beta$ of the four-site TFIM with $J = 3, h = 1$ versus inverse temperature $\beta$ using full and stochastic trace evaluation. QITE is performed with recompiled $D = 2$ unitaries with a time step of $\Delta\tau = 0.05$. Results of stochastic trace evaluation are shown with number of samples $n_{\text{samples}}$ set to 10 and 20. Inset shows the running average of $\expval{E}_\beta$ versus $n_{\text{samples}}$ using stochastic trace evaluation at $\beta = 0.2$ (red symbols), with full trace evaluation (blue symbols) and exact results (black solid line) plotted as constant values. Stochastic trace evaluation with 10 samples is already sufficient to reproduce the results from full trace evaluation across a wide range of $\beta$.}
\label{fig:stochastic_trace}
\end{figure}

To explore the scalability of our method, we compare stochastic trace evaluation with full trace evaluation in calculating the finite-temperature energy of the four-site TFIM. Stochastic trace evaluation is performed by uniformly selecting initial states in the full trace evaluation result with recompiled $D = 2$ QITE unitaries. In \cref{fig:stochastic_trace}, we plot the stochastic trace evaluation results with 10 and 20 samples along with the full trace evaluation and exact results; the inset shows the running average of $\expval{E}_\beta$ versus number of samples $n_{\text{samples}}$. As can be seen from the figure, random sampling with 10 samples already reproduced the results from full sampling on all 16 initial states, indicating that using scalable sampling schemes is a promising approach to studying larger systems.

\begin{figure*}[h]
\centering{
\phantomsubcaption\label{fig:4-site_corrfunc}
\phantomsubcaption\label{fig:4-site_spec}
\includegraphics[width=0.98\textwidth]{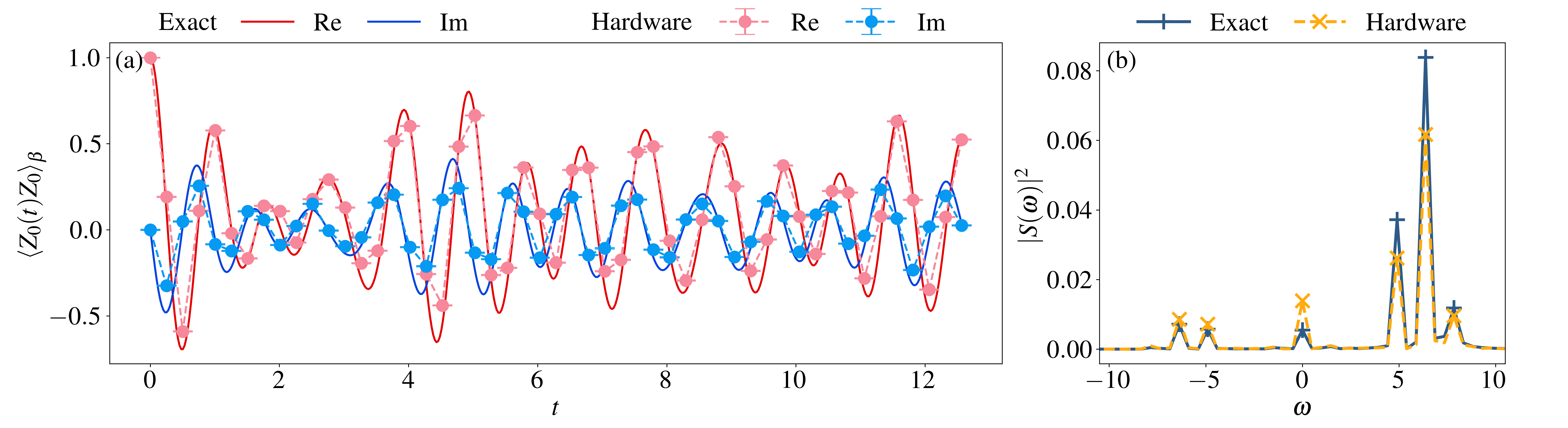}}
\caption{Finite-temperature dynamical properties of the four-site TFIM with $J = 3, h = 1$ at $\beta = 0.2$. QITE is performed with a time step of $\Delta\tau = 0.05$ and recompiled $D = 2$ unitaries. (a) Real and imaginary parts of the finite-temperature dynamical correlation function $\expval{Z_0(t)Z_0}_\beta$ versus real time $t$. Raw hardware data are post-processed by phase-and-scale correction. (b) Finite-temperature excitation spectra obtained by Fourier transform of exact and phase-and-scale-corrected hardware $\expval{Z_0(t)Z_0}_\beta$ at the same points in real time. The hardware $\expval{Z_0(t)Z_0}_\beta$ and excitation spectrum after phase-and-scale correction are in good agreement with the exact results.}
\label{fig:4-site_dynamical}
\end{figure*}

Finally, in \cref{fig:4-site_dynamical} we show the dynamical properties of the four-site TFIM with $J = 3, h = 1$ at $\beta = 0.2$. The calculation is implemented by recompiling $D = 2$ QITE unitaries with three gate rounds and real time propagation with five gate rounds. Figure \ref{fig:4-site_corrfunc} shows the real and imaginary parts of $\expval{Z_0(t)Z_0}_\beta$ after phase-and-scale correction. With this correction, both the real and the imaginary parts show good agreement with the exact result. Figure~\ref{fig:4-site_spec} shows the excitation spectra obtained by Fourier transforming the exact and phase-and-scale-corrected hardware $\expval{Z_0(t)Z_0}_\beta$ at the same points in real time. The excitation spectrum from hardware data accurately reproduces not only the frequencies $\omega = 0, \pm 4.90, \pm 6.37, 7.84$ but also the peak amplitudes.
 
The favorable agreement of the hardware $\expval{Z_0(t)Z_0}_\beta$ with the exact result is in contrast with the deviation of finite-temperature static observables from the exact results in \cref{fig:4-site_static}. In fact, the raw hardware $\expval{Z_0(t)Z_0}_\beta$ at $t = 0$ is 
\noindent $0.821 + 0.397i$, which is far from the exact value 1, indicating that phase-and-scale correction has a significant effect in correcting raw hardware data. Even though phase does not enter the static observables we computed, lack of a scale correction scheme for the static observables may explain their large deviation from the exact values compared to dynamical observables. More-

\onecolumngrid
\vspace{1em}

\twocolumngrid

\noindent over, even though the recompiled circuit in \cref{fig:4-site_dynamical} includes up to 11 gate rounds with the QITE and real time evolution gates combined, the ancilla is initialized after the QITE circuit and hence only experiences 5 gate rounds prior to measurement. The relatively shallow circuit applied to the ancilla may be another reason for the good performance of the quantum device for calculating the finite-temperature dynamical observable $\expval{Z_0(t)Z_0}_\beta$.

\section{Conclusion and Outlook} \label{sec:conclusion}

Our work demonstrates that finite-temperature physics of quantum many-body systems is accessible with near-term quantum hardware. With methods to reduce required quantum sources and mitigate errors in raw hardware data, QITE enables the practical calculation of finite-temperature energies, static and dynamical correlation functions, and spectral densities of excitations.

On two sites, static and dynamical observables for a wide range of temperatures are accurately captured by quantum hardware. An important factor underlying this accuracy is the constant depth of the circuit in both QITE and real time evolution. Constant depth in QITE allowed us to extend QITE-based finite-temperature calculations from a single site \cite{motta_2019}; constant depth in real time evolution allowed us to reproduce exact finite-temperature dynamical correlation functions on quantum hardware without phase-and-scale correction as compared to previous studies \cite{chiesa_2019,francis_2020}.

On four sites, finite-temperature static observables calculated on quantum hardware with circuit recompilation are in reasonable agreement with exact results at $\beta \leq 0.5$. We were also able to accurately reproduce the finite-temperature dynamical correlation function using phase-and-scale correction at a high temperature $\beta \sim 0.2$. However, accurate determination of observables at lower temperatures still appears challenging using the current recompilation scheme where the QITE unitaries are recompiled separately at each imaginary time step. Therefore, treating larger systems and lower temperatures will require additional reduction of circuit depth such as recompilation with merged imaginary time steps \cite{gomes_2020} or lower error rates on quantum devices either from efficient error mitigation for imaginary time or from improvements in hardware.

Scalable methods need to be employed for systems of larger size. For this reason, we examined how stochastic trace evaluation performs in calculating finite-temperature observables compared to full trace evaluation. We found that on four sites stochastic trace evaluation reproduced full trace evaluation results accurately in the temperature regime we studied. Compared to the previously proposed QMETTS algorithm, stochastic trace evaluation has zero autocorrelation time. A detailed comparison of QITE-based computation of finite-temperature observables with different sampling schemes is certainly a topic worth exploring. Furthermore, with the availability of more qubits \cite{ibmq_falcon,google_supremacy_2019}, trading increased computational time due to sampling for an increased number of qubits via constructing density matrix purification states \cite{martyn_2019,wu_2019} may be another feasible direction for studying finite-temperature physics on near-term quantum hardware.

\section*{Acknowledgements}
The authors thank Yi Wang, Xiuqi Ma, Sarah Sheldon and Tanvi Gujarati for helpful discussions. S.S., A.T.K.T., and A.J.M. are supported by NSF Grant Number 1839204. R.N.T. and G.K.C. are supported by the US Department of Energy, Office of Science, Grant Number 19374. S.S. acknowledges Jody Burks and Gavin Jones for helping with access to IBM Quantum devices.

\appendix

\section{\uppercase{Proof of Pauli String Reduction by $\Z_2$ Symmetries}}
\label{app:proof_of_pauli_string_reduction}
In \cref{sec:pauli_string_reduction} we introduced a scheme to reduce Pauli strings in the QITE unitaries by $\Z_2$ symmetries. We mentioned that rather than impose $\Z_2$ symmetries in choosing the Pauli strings in the QITE unitries, the original QITE algorithm subsumes the preservation of $\Z_2$ symmetries. We now restate the proposition and present a proof that derives directly from the QITE linear systems in \cref{eq:Ax=b}. 

\begin{proposition} \label{proposition}
Suppose QITE is applied to approximate the imaginary time propagator $e^{-\Delta\tau\hat{H}[l]}$ on the state $\ket{\Psi}$. If there exists a stabilizer $\mathcal{S}$ such that every element of $\mathcal{S}$ commutes with $\hat{H}[l]$ and $\ket{\Psi}\in V_\mathcal{S}$, then\\
(a) The action of $e^{-i\Delta\tau\hat{G}[l]}$ on $\ket{\Psi}$ with $\sigma_{\bm{\mu}}\in \mathcal{P}_{\hat{H}[l]}$ is equivalent to the action with $\sigma_{\bm{\mu}}\in \mathcal{P}_{\hat{H}[l]}\cap \mathcal{N}(\mathcal{S})/\mathcal{S}$,\\
(b) $e^{-i\Delta\tau\hat{G}[l]} \ket{\Psi} \in V_\mathcal{S}$.
\end{proposition}

\begin{proof}
Pick $\sigma_{\bm{\mu}} \notin \mathcal{N}(\mathcal{S})$.
Since $e^{-\Delta\tau\hat{H}[l]}$ commutes with elements of $\mathcal{S}$ and $\ket{\Psi}\in V_\mathcal{S}$, for any $s\in \mathcal{S}$ we have $\bra{\Psi}e^{-\Delta\tau\hat{H}[l]}\sigma_{\bm{\mu}}s\ket{\Psi}$ = $-\bra{\Psi}s\,e^{-\Delta\tau\hat{H}[l]}\sigma_{\bm{\mu}}\ket{\Psi}$, which implies $\bra{\Psi}e^{-\Delta\tau\hat{H}[l]}\sigma_{\bm{\mu}}\ket{\Psi} = 0$. Hence
\begin{align}
b[l]_{\bm{\mu}} &= \frac{\Im \langle \Psi | e^{-\Delta\tau\hat{H}[l]} \sigma_{\bm{\mu}} | \Psi\rangle}{\Delta\tau c[l]^{1/2}} = 0.
\end{align}

\noindent Now fix the column index $\bm{\nu}$ such that $\sigma_{\bm{\nu}}\in \mathcal{N}(\mathcal{S})$, then for any $s\in \mathcal{S}$, $\bra{\Psi} \sigma_{\bm{\mu}} \sigma_{\bm{\nu}} s\ket{\Psi} = - \bra{\Psi}s\,\sigma_{\bm{\mu}} \sigma_{\bm{\nu}} \ket{\Psi}$, which implies $\bra{\Psi} \sigma_{\bm{\mu}} \sigma_{\bm{\nu}} \ket{\Psi} = 0$. Hence
\begin{align}
A_{\bm{\mu}\bm{\nu}} = \Re(\bra{\Psi} \sigma_{\bm{\mu}}\sigma_{\bm{\nu}}\ket{\Psi}) = 0
\end{align}

\noindent Since $\bm{A}$ is Hermitian and real, $A_{\bm{\nu}\bm{\mu}} = A_{\bm{\mu}\bm{\nu}}^* = A_{\bm{\mu}\bm{\nu}} = 0$. Thus the linear system has the block-diagonal form
\begin{align}
\begin{pmatrix}
\bm{A[l]'} & \bm{0} \\
\bm{0} & \bm{A[l]''}
\end{pmatrix}
\begin{pmatrix}
\bm{x[l]'} \\
\bm{x[l]''}
\end{pmatrix}
=
\begin{pmatrix}
\bm{b[l]'} \\
\bm{0}
\end{pmatrix},
\end{align}
\noindent where the quantities with single primes are indexed by $\bm{\mu}$ such that $\sigma_{\bm{\mu}} \in \mathcal{N}(\mathcal{S})$ and those with double primes are indexed by $\bm{\mu}$ such that $\sigma_{\bm{\mu}} \notin \mathcal{N}(\mathcal{S})$. By setting $\bm{x[l]''}$ to $\bm{0}$, the linear system is reduced to $\bm{A[l]'}\bm{x[l]'} = \bm{b[l]'}$.

To show that the set of $\sigma_{\bm{\mu}}$ can be reduced from $\mathcal{N}(\mathcal{S})$ to $\mathcal{N}(\mathcal{S})/\mathcal{S}$, suppose $\sigma_{\bm{\mu}}$ and $\sigma_{\bm{\mu'}}$ belong to the same coset in $\mathcal{N}(\mathcal{S})/\mathcal{S}$, then $\sigma_{\bm{\mu}'} = \pm\sigma_{\bm{\mu}}s$ for some $s\in \mathcal{S}$. In the QITE unitary $e^{-i\Delta\tau\hat{G}[l]} = \sum_{k=0}^\infty (-i\Delta\tau)^k (\sum_{\bm{\mu}} x[l]_{\bm{\mu}} \sigma_{\bm{\mu}})^k$, each term in the sum is a power of $-i\Delta\tau$ times a product of the form $\prod_{\bm{\nu}}(x[l]_{\bm{\nu}}\sigma_{\bm{\nu}})$. If a product term contains $x[l]_{\bm{\mu'}}\sigma_{\bm{\mu'}}$, the action of this term on $\ket{\Psi}$ is proportional to
\begin{align}
\left(\prod_{\bm{\nu''}}x[l]_{\bm{\nu''}}\sigma_{\bm{\nu''}}\right)
(x[l]_{\bm{\mu'}}\sigma_{\bm{\mu'}})
\left(\prod_{\bm{\nu'}}x[l]_{\bm{\nu'}}\sigma_{\bm{\nu'}}\right)\ket{\Psi}
\label{eq:one_term_on_psi}
\end{align}
\noindent In the product over $\bm{\nu'}$, each $\sigma_{\bm{\nu'}}\in \mathcal{N}(\mathcal{S})$, so $\prod_{\bm{\nu'}}(x[l]_{\bm{\nu'}}\sigma_{\bm{\nu'}})\ket{\Psi}\in V_\mathcal{S}$. Then \cref{eq:one_term_on_psi} is equivalent to
\begin{align}
\left(\prod_{\bm{\nu''}}x[l]_{\bm{\nu''}}\sigma_{\bm{\nu''}}\right)
(\pm x[l]_{\bm{\mu'}}\sigma_{\bm{\mu}})
\left(\prod_{\bm{\nu'}}x[l]_{\bm{\nu'}}\sigma_{\bm{\nu'}}\right)\ket{\Psi}
\end{align}
Since this applies to every pair of Pauli strings in the same coset, $\hat{G}[l]$ can be written as
\begin{align}
\hat{G}[l] = \sum_{\bm{\mu}} \widetilde{x[l]}_{\bm{\mu}} \sigma_{\bm{\mu}},
\end{align}
\noindent where $\bm{\mu}$ is chosen such that $\sigma_{\bm{\mu}} \in \mathcal{P}_{\hat{H}[l]} \cap \mathcal{N}(\mathcal{S}) / \mathcal{S}$, $\widetilde{x[l]}_{\bm{\mu}} = \sum_{\bm{\mu'}} \eta_{\bm{\mu'}} x[l]_{\bm{\mu'}}, \eta_{\bm{\mu'}} = \pm 1$ and $\bm{\mu'}$ is chosen such that $\sigma_{\bm{\mu'}} \in \sigma_{\bm{\mu}} \mathcal{S}$.

Since all Pauli strings on the exponent of $e^{-i\Delta\tau\hat{G}[l]}$ commute with elements of $\mathcal{S}$, $e^{-i\Delta\tau\hat{G}[l]}$ commutes with elements of $\mathcal{S}$ and hence $e^{-i\Delta\tau\hat{G}[l]} \ket{\Psi}\in V_\mathcal{S}$.
\end{proof}

Our Pauli string reduction scheme is related to the qubit encoding scheme that removes redundant qubits by exploiting $\Z_2$ symmetries reported in Ref.~\cite{bravyi_2017}. In the qubit encoding scheme, a Hamiltonian over some number of qubits is transformed to another Hamiltonian over a smaller number of qubits by a series of Clifford gates. Our Pauli string reduction scheme coincides with the qubit encoding scheme when the domain size $D$ equals the total number of qubits $N$, in the sense that the reduced set of Pauli strings in our scheme exactly corresponds to all Pauli strings in the encoded Hamiltonian with redundant qubits removed in the qubit encoding scheme.

However, because the weight of a Pauli string can change during the Clifford transformation, the two schemes differ when $D < N$. On the one hand, some Pauli strings can decrease in weight after encoding. If we include all Pauli strings with domain size $D$ in the encoded Hamiltonian, these Pauli strings might include those with domain size $D' > D$ in the original Hamiltonian, thus increasing the total number of Pauli strings. On the other hand, some Pauli strings can increase in weight after encoding and result in an increased cost of the QITE algorithm. As an example, consider performing QITE on a Hamiltonian with periodic boundary condition and the $\Z_2$ symmetry $Z_0Z_1Z_2Z_3$. One of the $D = 2$ Pauli strings is $X_0Y_3$. In the qubit encoding scheme, the symmetry operator $Z_0Z_1Z_2Z_3$ is transformed to $Z_3$ so that qubit 3 can be eliminated, but the weight-two Pauli string $X_0Y_3$ is transformed to the higher-weight Pauli string $X_0X_1Y_2$, thus requiring a larger QITE domain and increasing the overall cost of the algorithm. Therefore, in the present work, we use $\Z_2$ symmetries to reduce the number of Pauli strings in the QITE unitaries rather than eliminate redundant qubits.

\section{\uppercase{Error bars in trace evaluation}}
\label{app:error_bars}

We describe calculation of error bars of finite-temperature observables in full and stochastic trace evaluation. Here we use $\E(Q)$ and $\Var(Q)$ to denote the mean and variance of a quantity $Q$. The error in $Q$ is the square root of its variance.

A finite-temperature observable $\langle \hat{O}\rangle_\beta \equiv O$ has the expression $O = \sum_i P_i O_i / \sum_i P_i$, where $P_i = || \ket{\Psi_i(\beta/2)} ||^2$ is the (unnormalized) probability and $O_i = \langle \Psi_i(\beta/2) | \hat{O} | \Psi_i(\beta/2)\rangle$ is the expectation value of the observable after imaginary time evolution on the $i$th basis state $\ket{\Psi_i}$. On quantum computers, each probability $P_i$ is built up from the energy expectation value at each imaginary time step:
\begin{align}
P_i = \prod_{k=0}^{n_{\beta/2}-1} e^{-2\Delta\tau E_{i, k}},
\end{align}
\noindent where $E_{i, k} = \langle\Phi_i(k\Delta\tau/2) | \hat{H} | \Phi_i(k\Delta\tau/2)\rangle$ and $n_{\beta/2} = \beta/2\Delta\tau$; each $O_i$ is the expectation value of the observable on the QITE-evolved state $\ket{\Phi_i(\beta/2)}$. Note that here both the exact imaginary-time-evolved state $\ket{\Phi_i(\beta/2)}$ and the QITE-evolved state $\ket{\Psi_i(\beta/2)}$ are consistent with the definitions in \cref{sec:finite-temperature}.

In full trace evaluation $O$ is regarded a function of the $P_i$ and $O_i$, which are random variables because of measurement sampling on quantum computers. $\Var(O)$ can be evaluated by expanding $O$ to first order in all $P_i$ and $O_i$ and assuming all $P_i$ and $O_i$ are independent, which then gives
\begin{align}
\hspace{0em}
\Var(O) = \frac{\sum_{i=1}^{2^N} [\E(P_i)^2 \Var(O_i) + \left(\E(O_i) - \E(O)\right)^2\Var(P_i)]}{\left(\sum_{i=1}^{2^N}\E(P_i)\right)^2}
\end{align}

In stochastic trace evaluation, we need to consider initial state sampling on top of measurement sampling. Define the numerator $\mathsf{N} = n_{\text{samples}}^{-1} \sum_{i=1}^{n_{\text{samples}}} P_i O_i$ and the denominator $\mathsf{D} = n_{\text{samples}}^{-1} \sum_{i=1}^{n_{\text{samples}}} P_i$ so that $O = \mathsf{N} / \mathsf{D}$. By first-order expansion of $O$ in $\mathsf{N}$ and $\mathsf{D}$ and assuming $\mathsf{N}$ and $\mathsf{D}$ are independent,
\begin{align}
\Var(O) = \frac{\E(\mathsf{N})^2\Var(\mathsf{D}) + \E(\mathsf{D})^2\Var(\mathsf{N})}{\E(\mathsf{D})^4}.
\end{align}
Now expanding $\mathsf{N}$ and $\mathsf{D}$ to first order in all variables and assuming all variables are independent, we have
\begin{align}
\Var(\mathsf{N}) &= \frac{1}{n_{\text{samples}}^2} \sum_{i=1}^{n_{\text{samples}}} [(\E(P_i)\E(O_i) - \E(\mathsf{N}))^2 + \notag \\
&\quad + \E(P_i)^2\Var(O_i) + \E(O_i)^2\Var(P_i) ],\\
\Var(\mathsf{D}) &= \frac{1}{n_{\text{samples}}^2} \sum_{i=1}^{n_{\text{samples}}} [(\E(P_i) - \E(\mathsf{D}))^2 + \Var(P_i)].
\end{align}

\ \\

\bibliography{main}

\end{document}